\documentstyle[aps,prb,twocolumn]{revtex}

\begin{document}
\title{Multivalued dependence of the magnetoresistance on the quantized 
conductance in nanosize magnetic contacts}
\author{L.R. Tagirov$^1$, B.P. Vodopyanov$^{2,1}$, and K.B. Efetov$^{3,4}$}
\address{$^1$Kazan State University, 420008 Kazan, Russia\\
$^2$Kazan Physico-Technical Institute of RAS, 420029 Kazan, Russia\\
$^{3}$Theoretische Physik III, Ruhr-Universit\"{a}t Bochum, 44780 Bochum,
Germany \\
$^{4}$L.D. Landau Institute for Theoretical Physics, 117940 Moscow, Russia}
\date{\today{}}
\maketitle
\draft

\begin{abstract}
We calculate the quantized conductance of nanosize point contacts between
two ferromagnets for different mutual orientations of the magnetic moments.
It is found that the magnetoresistance MR is a multivalued function of the
quantized conductance at the parallel alignment of the magnetizations 
$\sigma ^{F}$. This leads us to the conclusion that experimentally observed
large fluctuations of MR versus $\sigma ^{F}$ are rather due to the
conductance quantization than to measurement errors or a poor
reproducibility of the results. Using the results of the calculations we are
able to understand experimental data obtained by Garc\'{i}a {\it et al }for
MR of the magnetic nanocontacts.
\end{abstract}

\pacs{PACS numbers: 74.80.Dm, 74.50.+r, 74.62.-c}

\section{Introduction}

Recently a giant magnetoresistance (GMR) exceeding 200\% was discovered by
Garc\'{i}a {\it et al} in Ni-Ni \cite{Garcia1} and Co-Co \cite{Garcia2}
point contacts at room temperature. Somewhat smaller ($\sim 30\%$ ) but also
very large magnetoresistance was observed in Fe-Fe point contacts. \cite
{Garcia3}. These experiments revealed large fluctuations in the measured
values of the magnetoresistance drawn versus the conductance at
ferromagnetic alignment of magnetizations in contacts, $\sigma ^{F}$
(F-conductance). For Ni-Ni and Co-Co contacts, the fluctuations are
especially large at $\sigma ^{F}$ of the order of several elementary
conductances $e^{2}/h$, which may indicate that the effect observed is
related to a conductance quantization.

The quantization of the conductance in magnetic nanosize contacts has been
observed experimentally in Refs. \onlinecite{Costa,Oshima,Ono}. 
Costa-Kr\"{a}mer \cite{Costa} and Oshima and Miyano \cite{Oshima} reported 
on an odd integer number $N$ of open conductance channels ($\sigma =N(e^{2}/h)$) 
in nickel point contacts at room temperature. Ono {\it et al} \cite{Ono}
presented an evidence of changing the conductance quantum from $2e^{2}/h$ to 
$e^{2}/h$ at room temperature in nickel nanocontacts of another morphology.
Imamura {\it et al} \cite{Imamura} and Zvezdin and Popkov \cite{Zvezdin}
have calculated the conductance of a point contact between two ferromagnets
and demonstrated the $e^{2}/h$ conductance quantization due to a
non-simultaneous opening of ``up'' and ``down'' spin-channels. Imamura {\it
et al} \cite{Imamura} also studied numerically the magnetoresistance as a
function of F-conductance $\sigma ^{F}$ and came to the conclusion that, in
the conductance quantization regime, the magnetoresistance oscillated as a
function of the conductance.

In this paper, we calculate the conductance and the magnetoresistance of
nanosize magnetic contacts in the regime of conductance quantization. We
found that, at low temperatures, the magnetoresistance is a multivalued
function of the conductance at the parallel alignment of magnetizations, 
$\sigma ^{F}$. In other words, in the regime of quantization, different
samples, having the same F-conductance $\sigma ^{F}$, may have different
magnetoresistances. The distribution of the magnetoresistance is extremely
broad for the first few open F-conductance channels. This leads us to the
conclusion that large data fluctuations observed in the experiments by
Garc\'{i}a {\it et al} \cite{Garcia1,Garcia2,Garcia3} may be a direct
consequence of the conduction quantization. This means that the data
fluctuations are inevitable for the magnetoresistance measurements in the
nanosize magnetic contacts and this effect should not be treated as being
due to experimental errors or a poor reproducibility of the measurements.

\section{Basic formulae for the conductance and the magnetoresistance}

In a recent paper \cite{Tagirov} we applied a quasiclassical (QC) method for
calculations of the conductance of point contacts between ferromagnetic
metals. We considered a model of two ferromagnetic, single domain half
spaces contacting each other through a circular hole of a radius $a$ in an
impenetrable membrane, separating the domains. At antiferromagnetically (AF)
aligned domains, a domain wall (DW) is created inside the constriction. We
argued that the giant magnetoresistance values were determined by
peculiarities of the carrier transmission through DW.

If the spin direction does not change when passing through DW, the carriers
are strongly reflected by the interface. This effect can easily be
understood because, in this situation, the electron moves effectively in a
step-like potential. Of course, this reflection is large if the change of
the magnetization in the constriction occurs at short distances. The above
scenario may be realized provided the DW width is small, $d_{W}<d_{s}$,
where $d_{s}=\min (\frac{v_{F}}{\omega _{Z}},v_{F}T_{1})$, $T_{1}$ is the
longitudinal relaxation rate time of the carriers magnetization, and $\omega
_{z}$ is the Zeeman precession frequency. \cite{Gregg} In this limit, the
carrier spin does not have enough time to follow the magnetization profile
in DW. The strong reflection on DW leads to the magnetoresistance of the
order of few hundred percents, if one uses reasonable values of spin
polarizations of the conduction band estimated from the experimental data of
Refs. \onlinecite{TedMes,Soulen,Buhrman,Nadgorny}.

In this paper, we use the model described above for the case when the
conductance of the constriction is quantized. The connecting hole is assumed
to have a cylindrical shape of arbitrary (but shorter than the mean free
path $l$) length $d$. In the case of F-alignment of the magnetizations, the
carriers move effectively in a constant potential. For the AF-aligned
domains, the carriers move in a potential corresponding to the magnetization
profile of the domain wall (Ref. \onlinecite{Bruno}, Fig. 2). The hole
connecting the two parts of the space plays the role of a filter selecting
only those incidence angles that are allowed by the energy and momentum
conservation. As the diameter of the hole is assumed to be very small, we
may use the ballistic-limit versions of Eqs. (14), (18) and (19) of our work 
\cite{Tagirov} to calculate the conductance of the constriction: 
\begin{eqnarray}
\sigma ^{F} &=&\sigma _{\uparrow \uparrow }+\sigma _{\downarrow \downarrow }
\nonumber \\
\ &=&\frac{e^{2}}{h}\widetilde{\sum_{m,n}}\left\{ D_{\uparrow \uparrow
}(x_{mn})+D_{\downarrow \downarrow }(x_{mn})\right\} ,  \label{eq1}
\end{eqnarray}
\begin{equation}
\sigma ^{AF}=\frac{2e^{2}}{h}\widetilde{\sum_{m,n}}D_{\uparrow \downarrow
}(x_{mn}).  \label{eq2}
\end{equation}
Similar formulae can also be obtained within the Landauer-B\"{u}ttiker
scattering formalism \cite{Landauer}. In the above expressions, $\sigma
^{F}(\sigma ^{AF})$ is the conductance at ferromagnetic (antiferromagnetic)
alignment of the domains, $\sigma _{\alpha \alpha }$ is the conductance for
the $\alpha $-th spin channel, and $x_{mn}=\cos \theta $ is the cosine of
the quasiparticle incidence angle, $\theta $, measured from the cylinder
axis direction, the allowed values of $\cos \theta $\ are defined by Eq. 
(\ref{eq4}). $D_{\alpha \beta }(x)$ is the quantum mechanical transmission
coefficient for the connecting hole. Calculation of this coefficient is
straightforward, but lengthy. It is presented in the Appendix, and an
explicit expression for $D_{\alpha \beta }\left( x\right) $ is given by Eqs.
(\ref{eq24})-(\ref{eq30}).

The conductance quantization is assumed to be due to the quantization of
transversal motion in the constriction. In the ballistic regime, when
disorder is neglected, the quantization of the transversal motion in the
hole imposes the following condition for the component $p_{\parallel }$ of
the quasiparticle momentum parallel to the interface 
\begin{equation}
p_{\parallel }=p_{F\alpha }\sin \theta =p_{mn}\equiv \hbar a^{-1}Z_{mn},
\label{eq3}
\end{equation}
where $p_{F\alpha }$ is the Fermi momentum for the $\alpha $-th spin
channel, $Z_{mn}$ is the $n$-th zero of the Bessel function ${\bf J}_{m}(x)$
(see Appendix) and $a$ is the radius of the hole. The assumption of
the ballistic motion is quite reasonable provided the size of the hole is
much smaller than the mean free path $l$. We assume everywhere in this paper
that the inequality $a\ll l$ is fulfilled.

Eq. (\ref{eq3}) is the {\em first basic} selection rule. Tilde in Eqs. (\ref
{eq1}) and (\ref{eq2}) means that the summations should be done over the
open conduction channels satisfying the condition: 
\begin{equation}
x_{mn}\equiv \cos \theta =\sqrt{1-(\hbar Z_{mn}/p_{F\alpha }a)^{2}}\leq 1.
\label{eq4}
\end{equation}

When the alignment of the magnetizations is ferromagnetic, the Fermi momenta
on both sides of the contact are equal to each other in the each spin
channel. The energy and momentum conservation is already taken into account
in Eq. (\ref{eq1}) (both the ingoing and outgoing quasiparticles have the
same Fermi energy and the specular character of the scattering follows
automatically).

At the antiferromagnetic alignment, the conservation of the momentum
parallel to the interface ($p_{\Vert }\equiv p_{F1\alpha }\sin \theta
_{1}=p_{F2\alpha }\sin \theta _{2}$ , where the subscript 1 or 2 labels
left- or right hand side of the contact, respectively) introduces the 
{\em additional} selection rule into Eq. (\ref{eq3}): 
\begin{equation}
p_{F\alpha }=\min (p_{Fj\uparrow },p_{Fj\downarrow }).  \label{eq5}
\end{equation}
This selection rule strictly holds when the spin of the electron does not
change during the electron flight through the DW. This situation is realized
in the model of quantum DW \cite{Imamura} and in the model of effectively
abrupt DW. \cite{Tagirov}

The magnetoresistance is defined as follows \cite{Footn1} 
\begin{equation}
MR=\frac{R^{AF}-R^{F}}{R^{F}}=\frac{\sigma ^{F}-\sigma ^{AF}}{\sigma ^{AF}}.
\label{eq6}
\end{equation}
The value of the magnetoresistance is sensitive to the profile of the DW and
can become very large for sharp changes of the magnetization. We give an
exact solution to a problem for the linear profile of magnetization in
DW, which approximates well the behavior of magnetization in a narrow
constriction.\cite{Bruno} The limiting case of an infinitely steep slope
corresponds effectively to the electron motion in a step-like potential, and
gives the maximum possible magnetoresistance. In principle, the solution of
the problem for other domain wall profiles can be found by perturbations to
our exact solution. However, if the thickness of DW becomes comparable with
the Fermi wave-length of the current curriers, then DW becomes effectively
sharp even for the classical hyperbolic tangent profile of the magnetization
in DW.\cite{Falicov}

\section{Results of magnetoresistance calculations}

In order to find the conductances and the magnetoresistance of the
constriction one should take zeros $Z_{mn}$ of the Bessel function, ${\bf J}
_{m}(Z_{mn})=0$, and use the constraint, Eq. (\ref{eq3}). Determining the
transmission coefficients $D_{\alpha \beta }\left( x\right) $ (see Appendix)
we substitute it into Eqs. (\ref{eq1}), (\ref{eq2}) and perform summation
over the open channels.

At the ferromagnetic alignment of the magnetizations the equality 
$p_{F\alpha }\equiv p_{F\uparrow }$ is fulfilled for the $\sigma _{\uparrow
\uparrow }$ contribution to the conductance $\sigma ^{F}$, Eq. (\ref{eq1}),
and $p_{F\alpha }\equiv p_{F\downarrow }$ for the $\sigma _{\downarrow
\downarrow }$ contribution. At the antiferromagnetic alignment the minority
Fermi momentum should be used instead of $p_{F\alpha }$ in Eqs. (\ref{eq3})
and (\ref{eq4}) to calculate the conductance $\sigma ^{AF}$, Eq. (\ref{eq2}
). The results are displayed on Figures 1 and 2. The parameter $\delta
=p_{F\downarrow }/p_{F\uparrow }\leq 1$ characterizes the conduction band
spin polarization and is important for discussion. One can see from the
calculations that the results depend on the absolute value of $p_{F\uparrow
} $ and, to be specific, we have chosen $p_{F\uparrow }=1$\AA $^{-1}$.

Fig.1 displays the results of the calculations for $\delta =0.7$. The panel
(a) shows the dependence of F- and AF-conductances on the channel radius.
The parameters $d$ and $\lambda =dp_{F\uparrow }\hbar ^{-1}$ are the length
and dimensionless length of the channel, respectively. The chosen value, 
$\lambda =10.0$, corresponds to the length $d=10$\AA . The panel (b) shows
the dependence of the magnetoresistance on the radius of the hole. The
panels (c) and (d) display the magnetoresistance against F-conductance for a
potential with a finite slope (c), and for a step-like (d) potential in the
hole. Physically, Fig.1 corresponds to the case, when the AF-alignment
conduction opens up in the interior part of the first F-conductance plateau.
It allows us to make the following conclusions:

1) the F-alignment conductance is spin dependent and the spin channels open
non-simultaneously (see panel (a)), thus resulting in $e^{2}/h$ quantization
of the conductance \cite{Imamura,Zvezdin};

2) finite magnetoresistance appears simultaneously with the first spin
''down'' and AF-conductance (panel (b) in correlation with panel(a));

3) the magnetoresistance has quasi-periodic oscillations as a function of
the hole radius (panel (b));

4) sudden jumps in the magnetoresistance followed by practically flat
plateaus appear at points where a new F-alignment spin ``up'' conductance
channel opens up. They persist until the spin ``down'' projection opens a
new channel (panel (b) in correlation with panel (a));

5) when increasing the hole radius (panel (b)) or the number of open
channels (panels (c) and (d)), the amplitude of oscillations and of
sub-steps of the magnetoresistance decreases and its asymptotic value (panel
(d)) is given by our quasiclassical theory; \cite{Tagirov}

6) the most intriguing finding is that the magnetoresistance versus the
F-alignment conductance is a multivalued function of F-conductance, $\sigma
^{F}$ (panels (c) and (d)).

The results 2)-6) are novel, and let us discuss them in detail. The result
2) demonstrates that the magnetoresistance has the sharp peak when the first
conduction channel opens up for the spin ``down'' electrons at F-alignment.
If the spin polarization of conduction band is such that the spin ``down''
conductance channel appears at the first spin ``up'' conductance plateau,
the MR peak should appear at the conductance corresponding to $N^{F}=2$ open
channels of F-conductance, $\sigma ^{F}=(e^{2}/h)N^{F}$. This is displayed
in our Fig.1, and the experimental results by Garc\'{i}a {\it et al} \cite
{Garcia1,Garcia2,Garcia3} for Ni-Ni and Fe-Fe point contacts clearly
demonstrate the same tendency.

The result 3) indicates, that consecutive maxima in magnetoresistance as a
function of the contact radius correspond to opening of the AF-conductance
channels.

The result 4) leads to a weakly disperse or even non-disperse behavior of
magnetoresistance at certain numbers of open F-alignment channels: 
$N^{F}=3,4,7,8,10,12,13...$ . The non-disperse behavior of MR is due to the
fact that the AF-conductance is practically independent of the contact
radius when a new F-conductance channel opens up (see panel (b)).

The result 5) shows that if conductance exceeds the value 10-15 (in the 
$e^{2}/h$ units), which corresponds to the values of 8-10\AA\ for the hole
radius, the magnetoresistance fluctuations become relatively small and its
mean value converges well to that obtained in the ballistic quasi-classical
regime.\cite{Tagirov}

The result 6) is crucial for the interpretation of the experimental data.
The panel (b) shows a very sharp peak between $a\sim 2.65$\AA\ and 
$a\sim 3.8$\AA\ .\ The decay of the peak persists until a new spin
``down'' conduction channel opens at the F-alignment. If we draw the peak
magnetoresistance versus the number $N^{F}$\ of open channels (panels (c)
and (d)) we see that all the points correspond to the single abscissa, 
$N^{F}=2.$ This means that the magnetoresistance is a {\em 
multivalued }function of the number of open conduction channels at the
F-alignment, provided the temperature effects and quenched disorder may be
neglected. The magnetoresistance does not oscillate as a function of the
conductance $\sigma ^{F}$, but there are distributions of MR at fixed values
of the F-alignment conductance, $\sigma ^{F}$. The origin of these
distributions is clarified by inspection of the panel (b) of Fig.1 and its
comparison with the panel (a): in spite of the fact that the actual radius
may vary in the range $2.65-3.8$ \AA ,\ it gives identical values of the
F-conductance, which are due to the quantization. At the same time, the
AF-conductance depends on the radius (area) of the connecting channel, and
this results in the different values of magnetoresistance. The multivalued
magnetoresistance as a function of $\sigma ^{F}$ we predict is simply a
consequence of the conductance quantization. This property survives for
every reasonable shape of the nanocontact, provided that conductance at the
ferromagnetic alignment of magnetizations is quantized (conductance steps
exist), and the domain wall in the constriction is effectively sharp. The
multivalued behavior leads to extremely large fluctuations in the measured
magnetoresistance data at the same F-conductance values. The density of
points is considerably larger at small values of the magnetoresistance, than
at larger ones. As a consequence of the decreasing density of the points,
large values of the magnetoresistance are much less probable than the small
ones. When observed experimentally, such a MR distribution should not be
interpreted as being due to a poor reliability and reproducibility of
experimental data. The giant data fluctuations are inevitable in the
magnetoresistance measurements on the quantum magnetic contacts.

Increasing the spin polarization of the conduction band (decreasing the
parameter $\delta $) we see from our model that opening of the spin ``down''
conductance channel moves towards the second step in F-conductance. Then,
weakly disperse MR distributions, which originate from the flat
magnetoresistance graph sections (panels(b) of Figs. 1 and 2), become
strongly disperse moving closer to the steps of the spin ``down''
F-conductance. Fig.2 is drawn using the parameter $\delta =0.55$, so that
the spin ``down'' conductance appears now at the second plateau of the spin
``up'' conductance at F-alignment (panel (a)). Obviously, the MR points
appear now at $N^{F}=3$ open conductance channels of F-alignment. The figure
reveals the seventh finding: the minimal number of open channels $N^{F}$\ at
which the magnetoresistance points appear allows us to estimate the lower
bound for the conduction band spin polarization parameter $\delta $.

A further increase of the conduction band polarization leads to the
following interesting behavior: a) MR points appear at $N^{F}=4$ and larger
numbers of the open F-conductance channels; b) the theory predicts a huge
enhancement of MR at high conduction band polarizations (small $\delta $). 
From our calculations we conclude that if nanosize point contacts
made of highly spin polarized metals ($\delta <0.4$: NiMnSb, LMSO, CrO$_{2}$ 
\cite{Soulen}) with the F-conductance in the range of 5-10 channels were
available experimentally, they would show MR of 1000\% and higher.

Our calculations, panels (c) of Figs. 1 and 2, show that the finite length
of the constriction does not influence qualitatively the results, which can
be deduced from the calculations for the step-like potential barrier
corresponding to DW. All above conclusions hold, but the magnitude and the
overall width of MR distributions decrease as compared to the results of
calculations for the model of the step-like potential describing DW in the
constriction (panels (d) of Figs. 1 and 2).

\section{Discussion of the experiments}

Our calculations show that, in the quantized conductance regime, the minimal
number of open F-conductance channels, at which the values of the
magnetoresistance appear, is determined by the conduction band polarization 
$\delta $. In Fig. 1 the AP-alignment conductance channel opens up at the
first plateau of the spin ``up'' conductance. Corresponding
magnetoresistance points appear at $N^{F}=2$ open F-conductance channels.
Our analysis shows that the threshold of the magnetoresistance rise moves
from $N^{F}=2$ to $N^{F}=3$ open F-conductance channels at $\delta \simeq
0.63$. The experiments by Garc\'{i}a {\it et al} on Ni-Ni contacts (Fig. 2b
of Ref. \onlinecite{Garcia1}, Fig. 1a of Ref. \onlinecite{Garcia2} and Fig.
2 of Ref. \onlinecite{Garcia3}) and on Fe-Fe contacts (Fig. 1a of Ref. 
\onlinecite{Garcia3}) clearly indicate that the MR points appear close to 
$N^{F}=2$ for the both materials. This means that $\delta $ for both Ni and
Fe is {\em larger} than 0.63 for our choice of the parameter $p_{F\uparrow
}=1$\AA $^{-1}$.

In contrast, the experimental data for Co-Co contacts, Fig. 2a of Ref.
\onlinecite{Garcia2}, and Fig. 2 of Ref. \onlinecite{Garcia3}, indicate that
MR appears at $N^{F}\simeq 3$ open channels, that is at the second plateau
of the spin ``up'' F-conductance (see panels (a, b) of Fig. 2). This
suggests that polarization of the conduction band in Co is higher (and 
$\delta $ is smaller) as compared with Ni and Fe, and allows us to estimate
the lower bound as $\delta ($Co$)\simeq 0.47-0.63$.

Additional information can be extracted from the distributions of the MR
points at small numbers of open F-conductance channels. We interpret these
distributions as manifestation of the multivalued behavior of the
magnetoresistance as a function of F-conductance. In Fig. 3, the calculated
values of MR are compared with the results of the experiments on Ni and Co
point contacts. Solid circles at every quantized value of the conductance
show the range of the magnetoresistance distributions calculated at $\lambda
=6.0$. In addition, some points close to the experimentally measured ones
are shown inside the regions. The maximum theoretically available MR values
for this length of the channel are $\sim 500$\% for Ni (at $N^{F}=2$) and 
$\sim 1600$\% for Co (at $N^{F}=3$).

Surprisingly, our simple model reproduces well the MR fluctuations and the
extreme MR values at $\delta ($Ni$)\simeq 0.64$ and $\delta ($Co$)\simeq
0.57 $ and small numbers of open conduction channels. A deviation of the
calculated MR from the experimental ones at $N^{F}\geq 6-8$ (in $e^{2}/h$
units) may be referred to various reasons:

1) when the diameter of the constriction becomes large the domain wall is no
longer effectively abrupt (independent of the actual shape), and the
magnetoresistance begins to drop very fast down to the values 2-11\% given
by the Levy-Zhang mechanism of scattering enhancement in the domain wall; 
\cite{Levy}

2) one or several impurities or lattice defects may be located just at the
constriction causing additional random deviations of conductance values from
integer numbers of $e^{2}/h$\ (see, for example, Refs.
\onlinecite{Costa,Chu,Mochal} and references therein);

3) the shape of the constriction may deviate substantially from the
cylindrical one. According to calculations by Torres et al \cite{Torres} for
variable cross-section constrictions with the hyperbolic geometry, the
conductance quantization steps survive at the opening solid angles up to 
90$^{\circ }$, at least for the small number of open conductance channels. 
This also leads to deviations of conductance values from integer numbers of
$e^{2}/h$;

4) the non-cylindrical cross-section of the connecting channel (it can be
verified for the elliptic and rectangular cross-sections) influences the
quantization conditions and, hence, the sequence of openings of the spin
channels. 

The latter reason may change the assignment of some open conduction channels
from the spin ``up'' quantization to the spin ``down'' one and {\it vice
versa}. It may influence the structure and the width of the MR distributions
at fixed values of the F-conductance $\sigma ^{F}$ in Fig. 3, but does not
destroy overall consistency of the theory with the experiment.

Thus, besides the conduction band polarization parameter $\delta $, a
contact size,\ a shape and a length of the channel determine the values of
the magnetoresistance. The real nanocontacts by Garc\'{i}a et al have been
made by pressing a sharpened ferromagnetic tip into another piece of a
ferromagnet. Every MR point has been measured for a particular contact with
individual shape, size and length of the constriction. That is why we
believe that overall agreement of the theory with the experiment is fairly
good. 

Thus, our analysis of Garc\'{i}a et al measurements \cite
{Garcia1,Garcia2,Garcia3} suggests that the conduction band polarization
parameters $\delta $ for different materials obey the following
inequalities: $\delta ($Fe$)\geq \delta ($Ni$)>\delta ($Co$).$ This means
that cobalt has the most polarized conduction band. On the other hand, one
can extract the information about the spin polarization of a ferromagnet's
conduction band from the Ferromagnet-Insulator-Superconductor (FIS)
tunneling spectroscopy and the F-S Andreev reflection spectroscopy. The FIS
tunneling data \cite{TedMes} provide the following estimates for the mean
values of the conduction band polarization parameter $\delta $: 0.63 for Ni;
0.48 for Co and 0.43 for Fe. The Andreev reflection spectroscopy \cite
{Soulen,Buhrman,Nadgorny} gives the mean values of $\delta $: 0.62 for Ni;
0.64 for Co; 0.62 for Fe - from Ref. \onlinecite{Soulen}; 0.72 for Ni and
0.68 for Co - from Ref. \onlinecite{Buhrman}; 0.6 for Ni and 0.62 for Fe -
from Ref. \onlinecite{Nadgorny}. So, our estimated values for $\delta $, 
$\delta ($Ni$)\simeq 0.64$ for Ni and $\delta ($Co$)\simeq 0.57$ for Co, are
rather close to those obtained from the tunnel and Andreev spectroscopies.
At the same time, these spectroscopies indicate that iron probably has the
highest polarization of the conduction band. This does not agree with the
conclusions obtained from measurements of the magnetic point contact
magnetoresistance.

Of course, complex band structures of the contacting metals in the Andreev
and the point contact measurements may affect the values of the conduction
band spin polarization obtained from these experiments. However, we would
like to stress here that this discrepancy may also be due to the character
of the electron transmission through the contact. In the point contact of
two ferromagnetic metals, an electron traverses the domain wall at the AF
alignment of magnetizations, which contrasts the tunneling in the tunnel and
Andreev spectroscopies. We believe, that regime of the spin conservation
during the flight through the constriction could not been satisfied in Fe.
Then, the effective domain width is large, $d_{w}($Fe$)\sim l_{s}($Fe$)$,
and the electron spin partially follows the domain wall profile. This
results in the appearance of the AF-conductance and magnetoresistance at
smaller numbers of open F-conductance channels ($N^{F}=2$), rather than at 
$N^{F}=3$ or $4$ as one might expect from our theory. This may be also the
reason why the magnitude of magnetoresistance is reduced in iron to $\sim
30\%$.

Let us say few words about influence of a disorder in the area of the
contact and of temperature effects on the magnetoresistance. Strong disorder
should be avoided in experiments, because it destroys quantization (see, for
example, Ref. \onlinecite{Mochal} and citations therein), and hence the huge
enhancement of magnetoresistance that we predict. One may estimate from Ref.
\onlinecite{Mochal} that typical range for fluctuations of the disorder
potential energy should be below 10\% of the Fermi energy. The experiments
by Garc\'{i}a et al have been made on pure metals at room temperature, so we
do not expect any effects that would result in the weak localization.\cite
{Tatara} If there is a single impurity just in the constriction, the
transmission coefficient changes \cite{Chu,Bagwell}, and this leads to
deviations of the F-conduction values from the integer numbers of $e^{2}/h$.
However, provided the condition $V_{i}/\varepsilon _{F}\ll 1$ is fulfilled,
where $V_{i}$ is the impurity potential, the effect of the impurity
scattering on the conductance is small, and our analysis for the ballistic
transmission should be qualitatively valid.

We expect that temperature effects are relatively small if the temperature
does not exceed the Fermi energy $\varepsilon _{F}$ and the Curie
temperature $T_{Curie}$, $k_{B}T\ll \varepsilon _{F},k_{B}T_{Curie}$.
Moreover, phonon and magnon assisted relaxation processes are quenched
because of a large, $1-3$ $eV$, exchange splitting of the conduction band.
The experimental observation of sharp conduction quantization steps in the
nickel nanosize contacts at room temperature \cite{Costa,Oshima,Ono}
confirms the above expectation.

From the above analysis we conclude that our theory is consistent with the
experimental data.\cite{Garcia1,Garcia2,Garcia3} Therefore, it is reasonable
to think that the origin of large fluctuations of the magnetoresistance as a
function of conductance $\sigma ^{F}$ at the ferromagnetic alignment is the
quantization of conductance, but not measurement errors or poor
reproducibility of the results. The smallest number of open F-conductance
channels, at which the magnetoresistance data appear, allowed us to estimate
the low bound of the spin polarization of the conduction band of a
ferromagnet. For a more detailed comparison of our theory with experiments
more experimental data points for the magnetoresistance, as well as more
experimentally determined or controlled parameters like $d,$ $T_{1},$ 
$\omega _{Z}$, and more information about the shape of the constriction are
needed. Besides the experiments by Garc\'{i}a et al, our theory has obvious
implications to future experiments with a nanocontact between two
ferromagnetic islands made of a short nanowire.

\section{Acknowledgments}

L.R.T. would like to thank Professor N.Garc\'\i a for discussion of the
experiments at MML01, Aachen. K.B.E. acknowledges the support by Deutsche
SFB 491 {\em Magnetische Heterostrukturen}. L.R.T. and B.P.V. acknowledge
the support by the Russian Foundation for Basic Research through the grant
No. 00-02-16328, and by NIOKR/AST through the grant No. 06-6.2-47/2001.

\section{Appendix: The transmission coefficient for the cylindrical channel}

In this Appendix we solve the quantum mechanical problem of a particle
motion in the cylindrical channel of the length $d$ and the radius $a$ and
find the exact coefficient of transmission through this channel. The
solution of the Schr\"{o}dinger equation in the cylindrical coordinates is
sought in the form 
\begin{equation}
\Psi (\rho ,x)=\Phi (x){\bf J}_{m}\left( \frac{Z_{mn}\rho }{a}\right)
e^{im\varphi },  \label{eq10}
\end{equation}
where $Z_{mn}$\ being the discrete set of zeros of the Bessel function: 
${\bf J}_{m}(Z_{mn})=0$, and the variable $x$ is chosen along the axis of the
cylinder. The boundary condition is chosen as: 
\begin{equation}
\Psi (a,x)=0.  \label{eq11}
\end{equation}
It results in the quantization of the transverse (with respect to the $x$
-axis) motion by the zeros of the Bessel function. The longitudinal motion
is described by the equation 
\begin{equation}
\hbar ^{2}\frac{\partial ^{2}\Phi }{\partial x^{2}}+\left(
p_{F0}^{2}-a^{-2}Z_{mn}^{2}\hbar ^{2}+2mU(x)\right) \Phi =0.  \label{eq18}
\end{equation}
In order to simplify the calculations we model the dependence of the
magnetization of the domain wall (Ref. \onlinecite{Bruno}, Fig.2) on the
coordinate $x$ perpendicular to the membrane by the following function 
\begin{equation}
U(x)=2Ix/d, -d/2<x<d/2.
\label{eq19}
\end{equation}
A general solution to Eq. (\ref{eq18}) can be written as, 
\begin{equation}
\Phi (x)=C_{1}{\rm Ai}(\xi )+C_{2}{\rm Bi}(\xi ),  \label{eq20}
\end{equation}
where ${\rm Ai}(\xi )$ and ${\rm Bi}(\xi )$ are the Airy functions, \cite
{Abramovitz} 
\begin{equation}
\xi (x)=\left( \frac{4mI\hbar }{d}\right) ^{-2/3}\left[ \frac{4mI}{d}
x-\left( p_{F0}^{2}-a^{-2}Z_{mn}^{2}\right) \right] .  \label{eq21}
\end{equation}
Introducing the spin ``up'' ($p_{F\uparrow })$ and spin ``down'' 
($p_{F\uparrow })$ Fermi momenta and using the relation, 
\begin{equation}
\frac{p_{F\uparrow }^{2}}{2m}-\frac{p_{F\downarrow }^{2}}{2m}=2I,
\label{eq22}
\end{equation}
we obtain 
\begin{eqnarray}
\xi (-\frac{d}{2}) &=&-\left( \frac{p_{F\uparrow }^{2}-p_{F\downarrow }^{2}}
{d\hbar ^{-1}}\right) ^{-2/3}\left[ p_{F\uparrow }^{2}-p_{\parallel }^{2}
\right] ,  \nonumber \\
\xi (\frac{d}{2}) &=&-\left( \frac{p_{F\uparrow }^{2}-p_{F\downarrow }^{2}}
{d\hbar ^{-1}}\right) ^{-2/3}\left[ p_{F\downarrow }^{2}-p_{\parallel }^{2}
\right] ,  \label{eq23}
\end{eqnarray}
where $p_{\parallel }=a^{-1}Z_{mn}\hbar $, is the parallel to the interface
(but transverse with respect to the $x$-axis of the cylinder) projection of
the momentum allowed by the quantization condition. Combining (\ref{eq20})
and (\ref{eq10}) we find the general solution of the Schr\"{o}dinger
equation for the particle moving in a cylinder.

To find the transmission coefficient through the cylinder, connecting two
bulk ferromagnetic metals, we match the wave functions and their derivatives
at the interfaces and relate the outgoing probability flux to the ingoing
one. As a result, the exact expression for the transmission coefficient
reads 
\begin{equation}
D=\frac{Num}{Denum},  \label{eq24}
\end{equation}
where 
\begin{equation}
Num=4p_{x1}p_{x2}\left[ \left| \gamma ^{2}\right| \phi _{1+}^{2}+\phi
_{2+}^{2}+(\gamma +\gamma ^{\ast })\phi _{1+}\phi _{2+}\right] ,
\label{eq25}
\end{equation}
\begin{eqnarray}
Denum &=&\left| \gamma ^{2}\right| \left[ p_{x1}^{2}\phi _{1-}^{2}+\left(
\phi _{2-}^{\prime }\right) ^{2}\right] +p_{x1}^{2}\phi _{2-}^{2}+\left(
\phi _{2-}^{\prime }\right) ^{2}  \nonumber \\
&&\ +(\gamma +\gamma ^{\ast })\left[ p_{x1}^{2}\phi _{1-}\phi _{2-}+\left(
\phi _{1-}^{\prime }\right) ^{2}\left( \phi _{2-}^{\prime }\right) ^{2}
\right]  \nonumber \\
&&\ +i(\gamma -\gamma ^{\ast })p_{x1}\left[ \phi _{1-}\phi _{2-}^{\prime
}-\phi _{2-}\phi _{1-}^{\prime }\right] .  \label{eq26}
\end{eqnarray}
In the above expressions 
\begin{eqnarray}
\phi _{1} &=&{\bf J}_{m}\left( \frac{Z_{mn}\rho }{a}\right) {\rm Ai}(\xi
)\cos (m\varphi +\gamma ),  \label{eq27} \\
\phi _{2} &=&{\bf J}_{m}\left( \frac{Z_{mn}\rho }{a}\right) {\rm Bi}(\xi
)\cos (m\varphi +\gamma ),  \label{eq28}
\end{eqnarray}
\begin{equation}
\phi _{i}^{\prime }=\frac{\partial \phi _{i}}{\partial x},\,\,\,\phi _{i\pm
}\equiv \phi _{i}\left( x=\pm \frac{d}{2}\right) ,  \label{eq29}
\end{equation}
\begin{equation}
\gamma =-\frac{\phi _{2+}^{\prime }-ip_{x2}\phi _{2+}}{\phi _{1+}^{\prime
}-ip_{x2}\phi _{1+}}.  \label{eq30}
\end{equation}
The projection of the momentum of incident particles on the $x$ -axis is 
\begin{equation}
p_{xi}=p_{Fi}\cos \theta _{i}.  \label{eq31}
\end{equation}
In the limit $d\rightarrow 0$ the expression for $D$ (\ref{eq24})
considerably simplifies and reduces to a familiar expression for the
transmission coefficient for scattering on the potential step, 
\begin{equation}
D_{step}=\frac{4p_{x1}p_{x2}}{\left( p_{x1}+p_{x2}\right) ^{2}},
\label{eq32}
\end{equation}
which has been used for checking the numerical calculations with $D$ 
(\ref{eq24}).

\begin{center}
{\bf Figure captions}
\end{center}

Fig. 1. The dependence of conductance (a), and magnetoresistance (b) on the
radius of the hole $a$. Panels (c) and (d) show dependencies of the
magnetoresistance on the number of the open conductance channels at the
F-alignment of the magnetizations: (c) for the potential described by Eq. 
(\ref{eq19}) (d) for the step-like potential. $\delta =0.7$ for all panels.

Fig. 2. The same as in Fig.1, but for $\delta =0.55$.

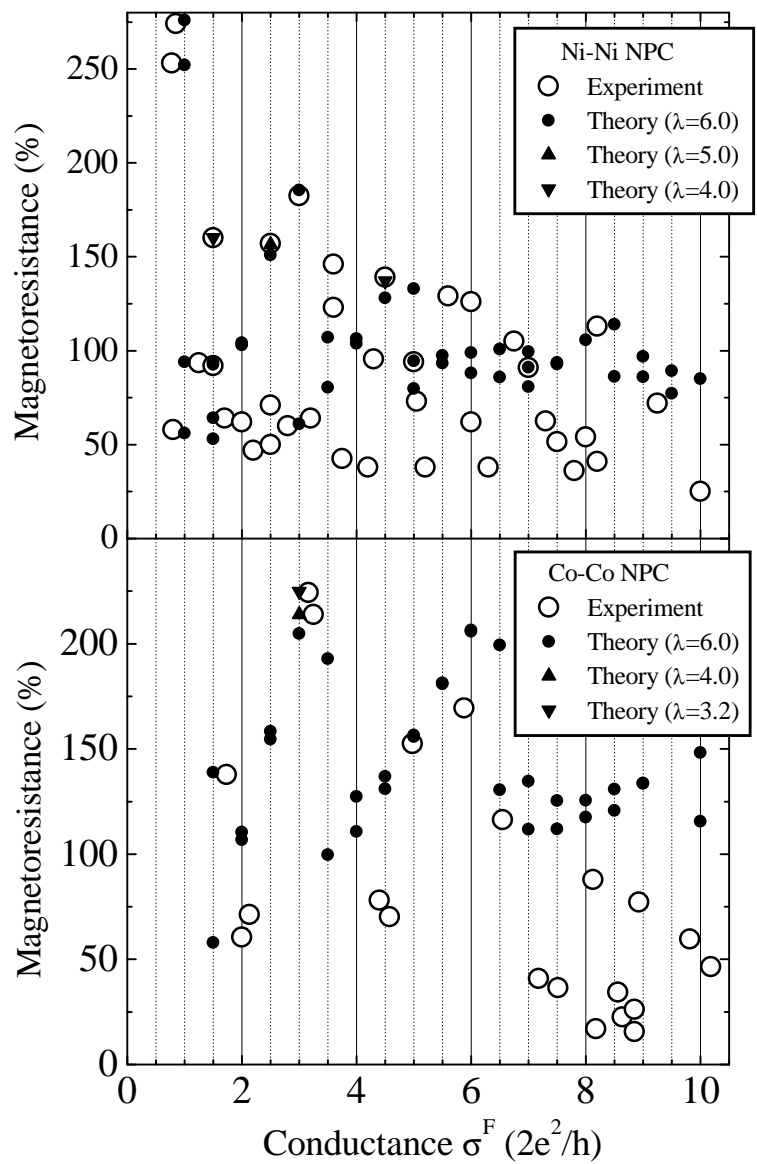
Fig. 3. Comparison between the theoretical and experimental values of the
magnetoresistance for Ni ($\delta =0.64$) and Co ($\delta =0.57$) nanosize
point contacts. The experimental data are taken from Ref.
\onlinecite{Garcia3}. For a discussion of the calculated MR values see the
text.


\begin{references}
\bibitem{Footn1}  Formally, the magnetoresistance defined by Eq. (6) may
diverge if the AF conductance $\sigma ^{AF}$ approaches zero, which would
mean that the nanocontact with one open channel of the F-conductance serves
as an ideal quantum spin-valve. However, this might be possible only if the
spin of conduction electrons conserves when the electrons traverse the
domain wall. The spin-flip process leads to a finite AF-conduction and
introduces an overall cutoff preventing the magnetoresistance to increase
infinitely. In this situation, there should be also a peak of the
magnetoresistance at one open channel of F-conductance. The reason why no
finite magnetoresistance has been observed in the experiment at one open
conduction channel for the F-alignment is unclear. Hereafter we will imply
that the conductance $\sigma ^{AF}$ in Eq. (6) is finite.
\end{references}
\end{document}